\documentclass[twocolumn,10pt,showpacs,preprintnumbers,nofootinbib]{revtex4}
\usepackage{url}
\usepackage{graphicx}
\usepackage{dcolumn}
\usepackage{bm}
\usepackage{amssymb}
\usepackage{amsmath}
\usepackage{epsfig}    
\usepackage{color}
\usepackage{slashed}
\usepackage{hyperref}
\usepackage{comment}
\usepackage[utf8]{inputenc}
\newcommand{\lam}{\lambda}

\begin{document}

\title{
Exploring the global symmetry structure of the Higgs potential\\
via same-sign pair production of 
charged Higgs bosons \\
}

\author{Masashi Aiko}
%\email{m-aikou@het.phys.sci.osaka-u.ac.jp}
\affiliation{Department of Physics, Osaka University, Toyonaka, Osaka 560-0043, Japan}

\author{Shinya Kanemura}
%\email{kanemu@het.phys.sci.osaka-u.ac.jp }
\affiliation{Department of Physics, Osaka University, Toyonaka, Osaka 560-0043, Japan}

\author{Kentarou Mawatari}
%\email{kentarou.mawatari@het.phys.sci.osaka-u.ac.jp }
\affiliation{Department of Physics, Osaka University, Toyonaka, Osaka 560-0043, Japan}

\preprint{OU-HET 995}

\begin{abstract}
\vspace*{5mm}
\begin{center}
{\bf Abstract}
\end{center}
\hspace*{12pt}
We propose a novel process where singly charged Higgs bosons are produced 
in a same-sign pair via vector boson fusion 
at hadron colliders in two Higgs doublet models. 
The process directly relates to the global symmetry structure of the Higgs potential. 
The produced charged Higgs bosons predominantly decay into a tau lepton and the neutrino 
or into a pair of top and bottom quarks, depending on the type of Yukawa interactions.
We evaluate the signal and the background for the both cases 
at the CERN Large Hadron Collider and future higher-energy hadron colliders. 
We find that the process can be feasible  
and useful to explore the nature of the Higgs potential. 
\pacs{\, 14.80.Cp, 13.85.-t %\hfill~~[\today] 
}
\end{abstract}

\maketitle

%[Introduction]

Since the discovery of the Higgs boson, 
it has been empirically confirmed that 
the idea of mass generation in the standard model (SM), 
which is based on the electroweak symmetry breaking (EWSB), is correct.  
Measured Higgs boson couplings with various SM particles have turned out to be consistent 
with the predictions of the SM under the experimental and theoretical uncertainties~\cite{Khachatryan:2016vau}, 
while no other new particle has been found up to now.    
The SM is a good description of Nature below the EWSB scale.
 
Although the Higgs boson was found, the structure of the Higgs sector remains unknown, 
and physics behind the EWSB is still mysterious. 
In the SM, the Higgs sector is assumed to be composed of an isospin doublet scalar field 
as a minimal realization. However, the Higgs sector causes the hierarchy problem, and  
new physics beyond the SM is expected to appear at TeV scales. 
In addition, there are phenomena which cannot be explained in the SM, 
such as dark matter, baryon asymmetry of the Universe and tiny neutrino masses. 
Mechanisms to explain these phenomena would also be related to the Higgs sector. 

Non-minimal Higgs sectors are often introduced in various new physics models 
motivated by the hierarchy problem or the above mentioned phenomena. 
They can be consistent with the current data like the SM. 
The multiplet structure is an important property; i.e., 
the number of additional scalars and their representations under the SM gauge symmetries~\cite{Gunion:1989we}.  
In addition, the global symmetry structure of the Higgs potential, 
whatever it is exact, approximate, softly broken or spontaneously broken, 
is also a key to physics beyond the SM. 
Therefore, in order to determine the nature of the EWSB 
and also to narrow dawn scenarios of physics beyond the SM, 
a detailed study of models with extended Higgs sectors is getting more and more important.    

In this Letter, we discuss an interesting new process 
to approach the global symmetry structure of the Higgs potential, 
which is pair production of same-sign charged Higgs bosons 
via vector boson fusion (VBF) at hadron colliders. 
Recently, ATLAS and CMS observed production of same-sign W boson pairs via VBF~\cite{Sirunyan:2017ret,Aaboud:2019nmv}. 
Having ongoing and future experiments, extended Higgs sectors can also be explored by using VBF processes.  

%[Model]

We consider the two Higgs doublet model (THDM) with isospin doublet scalar fields 
$\Phi_1$ and $\Phi_2$ with the hypercharge $1/2$.   
It is one of the well-motivated simple extensions among many candidates.  
The most general potential is given by
\begin{align}
&\hspace*{-0.3cm}V(\Phi_1,\Phi_2)
 = \sum_{i=1}^2 m_i^2\,|\Phi_i|^2 
   -(m_3^2\,\Phi_1^\dagger \Phi_2^{} + {\rm h.c.})\notag\\
 &\hspace*{-0.38cm}  + \sum_{i=1}^2 
   \frac{1}{2}\lambda_i\,|\Phi_i|^4 
  +\lambda_3\,|\Phi_1|^2|\Phi_2|^2+\lambda_4\,|\Phi_1^\dagger\Phi_2^{}|^2 \notag\\
 &  \hspace*{-0.38cm}  +\Big\{
  \frac{1}{2}\lambda_5\,(\Phi_1^\dagger\Phi_2^{})^2 
+\left( \lam_6 |\Phi_1|^2+\lam_7 |\Phi_2|^2 \right) \Phi_1^\dagger\Phi_2^{}
 +{\rm h.c.}\Big\}, \label{pot_thdm} 
\end{align}
where $m_{1,2}^2$ and $\lam_{1-4}$ are real, while $m_3^2$ and $\lam_{5-7}$ are complex.
Global symmetries of the potential in the THDM have been deeply investigated in 
Refs.~\cite{Deshpande:1977rw,Pomarol:1993mu,Haber:2010bw,Pilaftsis:2011ed}.

We here assume CP invariance, taking all parameters real. 
The fields $\Phi_{1,2}$ are parameterized as 
\begin{align}
\Phi_i=\left(\begin{array}{c}
\omega_i^+\\
\frac{1}{\sqrt{2}}(v_i+h_i+iz_i)
\end{array}\right), \quad i=1,2,  \notag
\end{align}
where $v_1$ and $v_2$ are vacuum expectation values, which satisfy   
$v=(v_1^2+v_2^2)^{1/2} = (\sqrt{2} G_F)^{-1/2} \simeq 246$~GeV with $G_F$ being the Fermi constant, 
and $\tan\beta=v_2/v_1$. 
Diagonalizing the mass matrix of CP-even neutral scalars by introducing the mixing angle $\alpha$, 
there are five mass eigenstates, two CP-even ($h$, $H$), CP-odd ($A$) and charged states ($H^\pm$). 
We here identify $h$ as the discovered Higgs boson.

%[Constraint on the Higgs potential]

Some important experimental constrains have been known on the Higgs potential~\cite{Tanabashi:2018oca}; 
a) suppressed flavor changing neutral current, and  
b) the rho parameter being close to unity. 
In addition, as new knowledge after the Higgs boson discovery, we have  
c) Higgs boson couplings being SM like~\cite{Khachatryan:2016vau}.    
These facts would suggest that at least approximately the Higgs potential respectively has 
a') a (softly-broken) discrete Z$_2$ symmetry ($\Phi_1 \to \Phi_1$ and $\Phi_2 \to - \Phi_2$)~\cite{Glashow:1976nt}, 
b') the custodial SU(2)$_{\rm V}$ symmetry~\cite{Sikivie:1980hm}, and  
c') the alignment~\cite{Gunion:2002zf,Dev:2014yca} 
when additional Higgs bosons are relatively light.
In the following discussion, we take these three conditions as guiding principles as the first approximation.

First, the condition a') requires $\lambda_6 = \lambda_7 =0$. 

Second, natural realization of the alignment c') 
would require that the off diagonal component of CP-even neutral scalars 
in the Higgs basis vanishes for all values of $\tan\beta$~\cite{Dev:2014yca}, which implies   
$\lambda_1=\lambda_2=\lambda_{3}+\lambda_4+\lambda_5$ and $\lambda_6=\lambda_7=0$. 
The quartic coupling part of the potential in Eq.~(\ref{pot_thdm}) is then written as 
\begin{align}
V_4 =& + \frac{1}{2} c_1 (|\Phi_1|^2+|\Phi_2|^2)^2 
+\frac{1}{2}c_2(|\Phi_1|^2-|\Phi_2|^2)^2 
\notag\\
&
+\frac{1}{2}c_3(\Phi_1^\dagger \Phi_2+\Phi_2^\dagger\Phi_1)^2 
+\frac{1}{2}c_4(\Phi_1^\dagger \Phi_2-\Phi_2^\dagger\Phi_1)^2, 
\label{eq:v4}
\end{align}
where 
$c_1 = \lambda_3 + (\lambda_4+\lambda_5)/2$, $c_2= c_3= (\lambda_4+\lambda_5)/2$, 
and $c_4=-(\lambda_4-\lambda_5)/2$.

Third, before the EWSB the term of $c_1$ in Eq.~(\ref{eq:v4}) is invariant under 
the maximal global symmetry O(8)%
\footnote{
For the entire Higgs sector of the THDM with the kinetic terms, 
SO(5) is maximal~\cite{Pilaftsis:2011ed}. 
}%
, 
and that of $c_2$ is  
invariant under O(4) $\times$ O(4)$'$ ($\subset$ O(8))~\cite{Deshpande:1977rw}.    
The term of either $c_3$ or $c_4$ is invariant under O(4), depending on the choice.  
After the EWSB, O(4) $\simeq$ SU(2)$_{\rm L}$ $\times$ SU(2)$_{\rm R}$ $\to$ SU(2)$_{\rm V}$. 
Two choices for b') the custodial SU(2)$_{\rm V}$ symmetry 
in $V_{4}$ are described by~\cite{Pomarol:1993mu,Haber:2010bw} 
\begin{align}
({\rm Case\; I})  & \quad c_4=(m_{H^\pm}^{2} - m_A^{2})/v^2 =0, 
\notag\\ 
({\rm Case\; II}) & \quad c_3=(m_{H}^{2} - m_{H^\pm}^{2})/v^2=0, \notag
\end{align}
where $m_\phi$ represents the mass of a field $\phi$. 

Finally, imposing simultaneously a') the discrete Z$_2$ symmetry, c') the natural alignment, 
and b') the custodial SU(2)$_{\rm V}$ symmetry 
to the quartic coupling part $V_4$ of the Higgs potential, we obtain  
\begin{align}
({\rm Case\; I})  & \quad c_1 =  m_h^2/v^2 - \eta, \; c_2 = c_3= \eta, \; c_4=0, 
\notag\\ 
({\rm Case\; II}) & \quad c_1 =   m_h^2/v^2,\; c_2=c_3=0,\; c_4=\eta,  \notag
\end{align}
where $m_h=125$GeV and $\eta =(m_H^{2} - m_A^{2})/v^2$. 
For both cases, $\eta$ is crucial for 
the global symmetry structure of $V_{4}$. 
If $\eta= 0$, $V_4$ respects O(8), otherwise O(4). 
Therefore, measuring $\eta$ by experiments is extremely important. 

%[New Process]

One of the ways to determine $\eta$ is to separately measure $m_H^{}$ and $m_A^{}$ 
by directly discovering $H$ and $A$.  
In this Letter, however, we discuss an interesting new process $W^\pm W^\pm \to H^\pm H^\pm$, 
whose amplitude is proportional to $\eta$, 
by which we can directly extract the global symmetry structure 
of the potential at collider experiments. 
In the alignment limit c'), which implies $\sin(\beta-\alpha)=1$, 
the helicity amplitudes are given by 
\begin{align}
{\cal M}_{\tilde{\lambda}_1 \tilde{\lambda}_2}(W^\pm W^\pm \to H^\pm H^\pm) 
= - 2 \eta \left(1 + {\cal O}(\eta)\right),  \notag
\end{align}
for all helicity set $(\tilde{\lambda}_1, \tilde{\lambda}_2)$ of the W bosons.    
This can be seen by the fact that, in the alignment limit, the diagrams of 
$H$ mediation and those of $A$ are destructive and exactly cancel for $m_H^{}=m_A^{}$.  
Null amplitude implies that $V_4$ respects a higher global symmetry 
than O(4), having $m_H^{} = m_A^{} = m_{H^\pm}^{}$~\footnote{
On the contrary, the amplitudes for the opposite-sign process  
$W^{+}W^{-}\to H^+H^-$ do not vanish for $m_H^{}=m_A^{}$.  
}.  
Therefore, at hadron colliders, we can directly measure $\eta$ by searching for the same sign pair production 
of charged Higgs bosons via W boson fusion, $pp \to W^{\pm\ast} W^{\pm\ast} jj \to H^\pm H^\pm jj$.
 
\begin{figure}
 \includegraphics[width=0.25\textwidth]{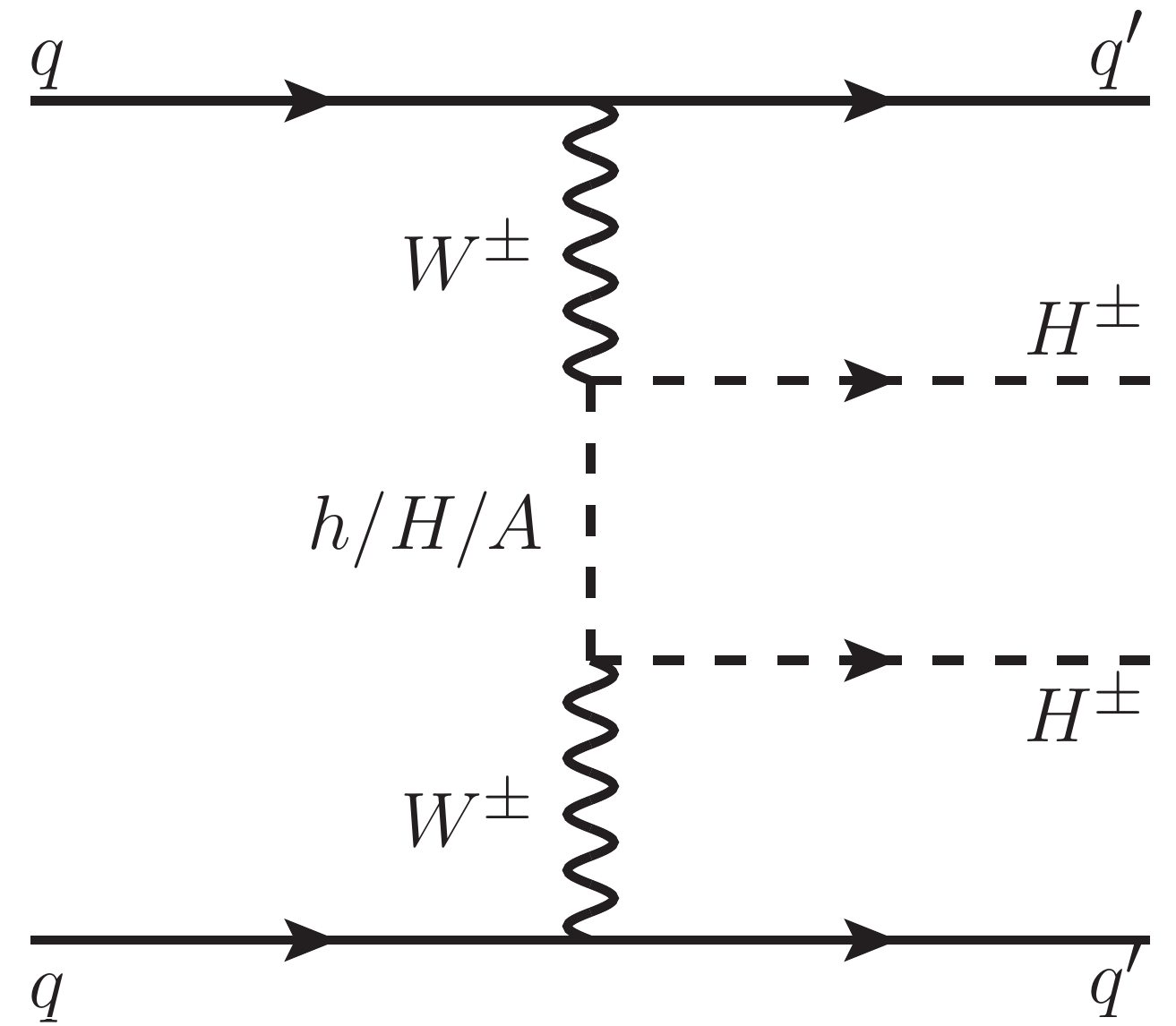}
 \caption{Feynman graphs (shown only the t-channel ones).
}
\label{fig:diagrams}
\end{figure} 

Relation between this process and the global symmetry of $V_4$ may also be understood partially as follows.  
In $V_4$, the term of $\omega^\pm \omega^\pm H^\mp H^\mp$ is given by $\eta$, where $\omega^\pm$ 
are the Nambu-Goldstone bosons to be absorbed by longitudinal components of $W^\pm$.
At high energies $\sqrt{s}\gg m_{W}^{}$, by the equivalence theorem~\cite{Cornwall:1974km}, we have  
\begin{align}
 {\cal M}_{00}(W^\pm W^\pm\to H^\pm H^\pm)  & \notag\\
 &\hspace{-3cm}\simeq {\cal M}(\omega^\pm \omega^\pm\to H^\pm H^\pm) \simeq -2\eta,  \notag
\end{align}
where ${\cal M}_{00}$ is the longitudinally polarized amplitude that 
dominates other polarizations in the high energy limit. 

Charged Higgs bosons have been searched at LEP~\cite{Abbiendi:2013hk} and 
the CERN Large Hadron Collider (LHC)~\cite{Aaboud:2018gjj}.
In Refs.~\cite{Akeroyd:2016ymd,Arbey:2017gmh,Arhrib:2018ewj}, the constraints on $m_{H^\pm}^{}$ in THDMs 
are summarized for each type of Yukawa interaction (Type-I, II, X or Y~\cite{Aoki:2009ha}). 
The mass below 160~GeV is excluded from $t\to H^\pm b$, $H^\pm\to\tau\nu$ searches
for $\tan\beta<8, 10, 4$ in the Type-I, -X and -Y THDM, respectively, while  
the same mass region is excluded irrespective of $\tan\beta$ in Type II.
Flavor experiments also give a strong constraint~\cite{Misiak:2017bgg}; e.g.,  
$m_{H^\pm}^{}>580$~GeV independently of $\tan\beta$ in Type-II and Type-Y.   
Including measurements of Higgs boson coupling strengths and flavor physics observables,
the constraints in THDMs are summarized in Ref.~\cite{Haller:2018nnx}.

%[Cross Sections]

\begin{figure*}
\includegraphics[height=0.8\columnwidth]{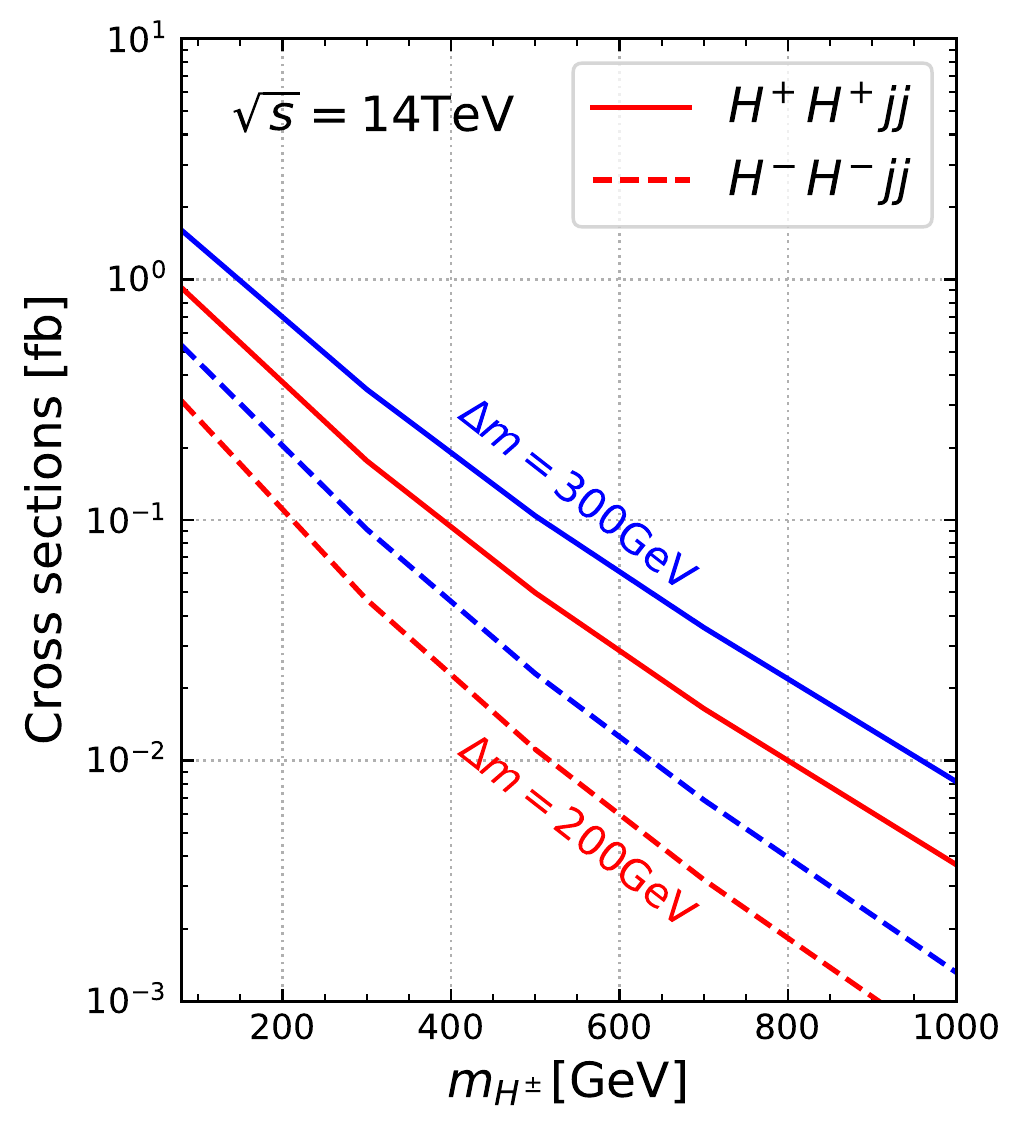}\hspace{12mm}
\includegraphics[height=0.8\columnwidth]{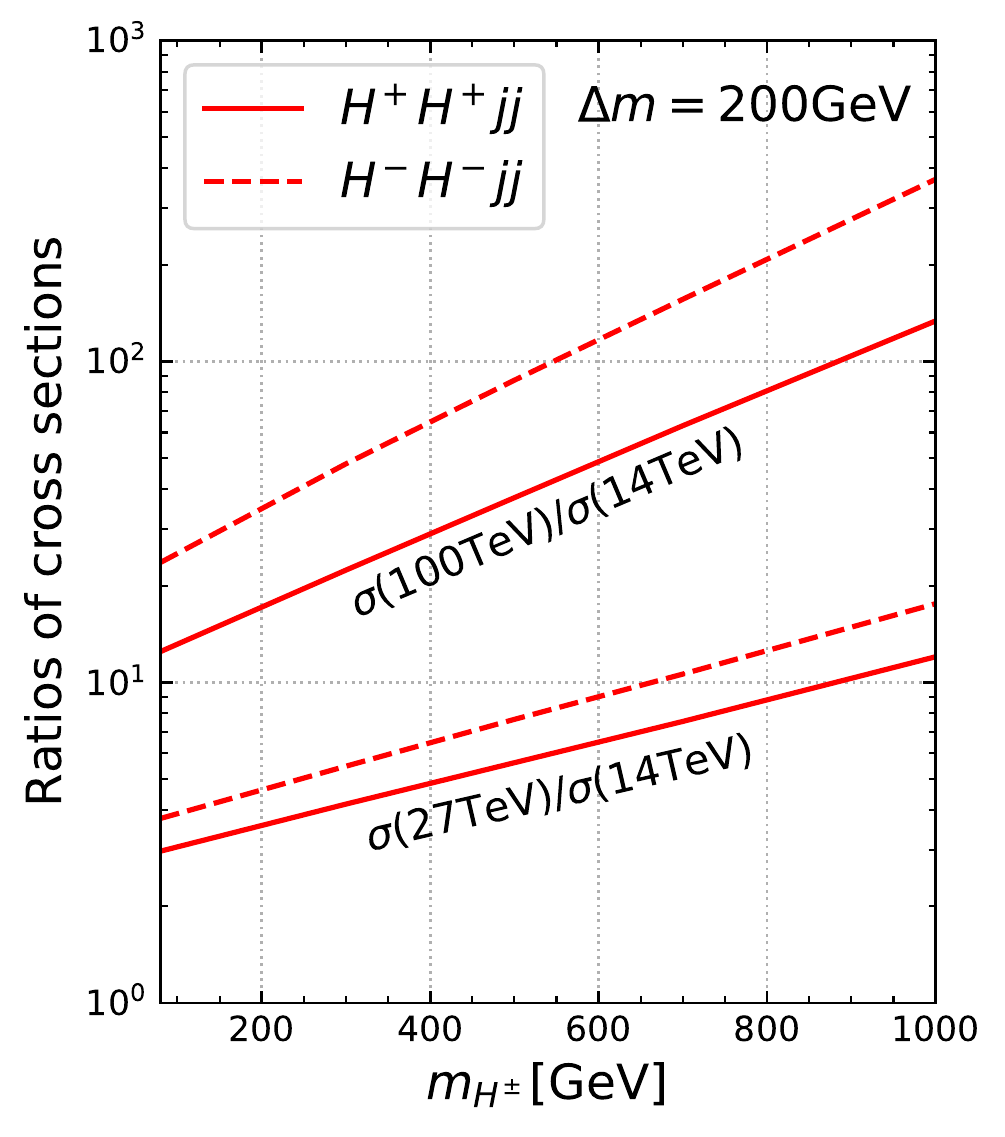}
 \caption{
 [Left] Cross sections for $pp\to H^\pm H^\pm jj$ at $\sqrt{s}=14$~TeV as a function of $m_{H^\pm}$  
 for $\Delta m (\equiv m_A^{}-m_H^{}) =200$ GeV and 300 GeV.
 [Right] Ratios of the cross sections evaluated at different $pp$ collision energies. }
\label{fig:xsec}
\end{figure*}
We now investigate the feasibility of the process 
$pp \to W^{\pm\ast} W^{\pm\ast} jj \to H^\pm H^\pm jj$
at the LHC as well as at future higher-energy hadron colliders.  
To evaluate cross sections and to generate events, 
we use the THDM UFO model file~\cite{Degrande:2014vpa}, 
and employ {\sc MadGraph5\_aMC@NLO}~\cite{Alwall:2014hca}.
The partonic total amplitude for $q_1^{}q_2^{}\to {W^\pm}^* {W^\pm}^* q'_1q'_2\to H^\pm H^\pm q'_1q'_2$ 
(see Fig.~\ref{fig:diagrams}) is calculated using Ref.~\cite{Hagiwara:2009wt}. 
Convoluting with the parton distribution functions, we obtain the total cross section shown in Fig.~\ref{fig:xsec}, where 
we required presence of at least two jets with the transverse momentum and the pseudorapidity as 
$p_T^j>30$~GeV and $|\eta^j|<5$.
In addition, we imposed an invariant mass cut and a rapidity separation cut on the two leading jets
as the VBF baseline selection, $m_{jj}>500~{\rm GeV}$ and $|\Delta\eta_{jj}|>2.5$.

In Fig.~\ref{fig:xsec}~(left) we show cross sections for $pp\to H^+H^+jj$ (solid) and $H^-H^-jj$ (dashed) at $\sqrt{s}=14$~TeV
as a function of $m_{H^\pm}^{}$ in the alignment limit with the custodial SU(2)$_{\rm V}$ symmetry of Case~II. 
The results are shown for $\Delta m (\equiv m_{A}^{} - m_H^{}) =200$~GeV and 300~GeV.  
As expected, the cross sections is larger for larger $\Delta m$, while it is zero for $\Delta m =0$ ($\eta=0$). 
In Fig.~\ref{fig:xsec}~(right), we show the ratios of the cross sections evaluated 
at $\sqrt{s}=27$~TeV and 100~TeV over those at 14~TeV for $\Delta m=200$~GeV.

%[Sensitivity]

We evaluate signal sensitivity for two decay modes of charged Higgs bosons, 
$H^\pm\to\tau\nu$ and $H^\pm\to tb$, by taking into account the SM background.
The $\tau\nu$ decay mode is dominant for $m_{H^\pm}^{} < m_t+m_b\simeq 180$~GeV or for $\tan\beta\gtrsim 20$ 
in Type-X, otherwise the $tb$ decay mode is dominant. 
Significance is defined by $\sqrt{2((s+b)\ln(1+s/b)-s)}$ with $s$ and $b$ respectively being 
event numbers of the signal and the background~\cite{Cowan:2010js}. 

First, signature for the signal $pp\to H^\pm H^\pm jj$ with $\ H^\pm\to\tau\nu$ is 
a same-sign tau-lepton pair with two forward jets.  
The main SM background is same-sign W boson pairs via VBF,
which was recently observed at the LHC~\cite{Sirunyan:2017ret,Aaboud:2019nmv}.
We estimate $s$ and $b$ with the integrated luminosity $L$ in the final state with 
two jets and two same-sign hadronic tau-leptons, $\tau^\pm_h\tau^\pm_hjj$, as
\begin{align}
  s = & L \;\sigma^{\rm VBF}_{pp\to H^+ H^+ jj} 
%  & \times 
   (B_{H^+\to\tau\nu})^2
   \epsilon_{\rm sel}^{\tau} (\epsilon_{\tau})^2, \notag\\
 b =& L\;\sigma^{\rm VBF}_{pp\to W^+ W^+ jj} \notag 
 (B_{W^+\to\tau\nu})^2
   \epsilon_{\rm sel}^{\tau} (\epsilon_{\tau})^2, 
\end{align}  
where $\sigma^{\rm VBF}$ is the cross section with the VBF selection. 
The signal cross sections are shown in Fig.~\ref{fig:xsec},
while the background rate is estimated at LO as 163 (568) fb at $\sqrt{s}=14$ (27) TeV. 
The decay branching ratio is taken as $B_{H^+ \to \tau\nu} =1$.  
We obtain the selection efficiencies $\epsilon_{\rm sel}^\tau$ by requiring the presence of two same-sign tau leptons with
$p_T^\tau>20$~GeV and $|\eta^\tau|<2.5$
as well as a larger rapidity separation between the leading jets. 
The efficiencies for the signal are 0.90--0.96 (0.80--0.93) for $m_{H^\pm}=200$--1000~GeV with $|\Delta\eta_{jj}|>2.5$ (4.5) at $\sqrt{s}=14$~TeV,
while the efficiency for the background is 0.57 (0.26). Those values are similar at $\sqrt{s}=27$~TeV. 
We also take into account $\epsilon_{\tau}=0.65\times 0.6$ as the hadronic-tau branching ratio~\cite{Tanabashi:2018oca} 
and the efficiency of the  
identification~\cite{Sirunyan:2018pgf}. 

\begin{figure*}
 \includegraphics[height=0.8\columnwidth]{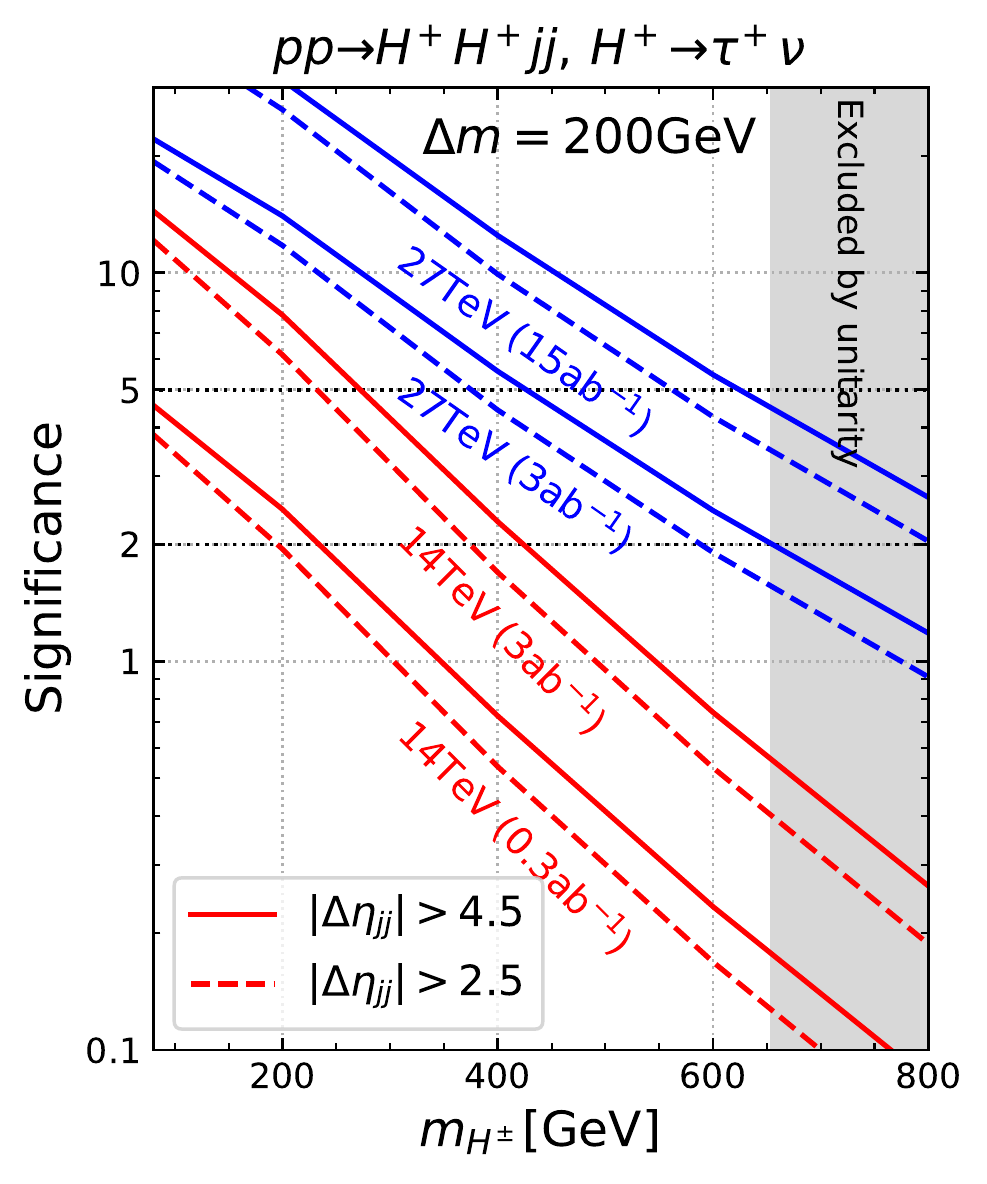}%
 \hspace{12mm}
 \includegraphics[height=0.8\columnwidth]{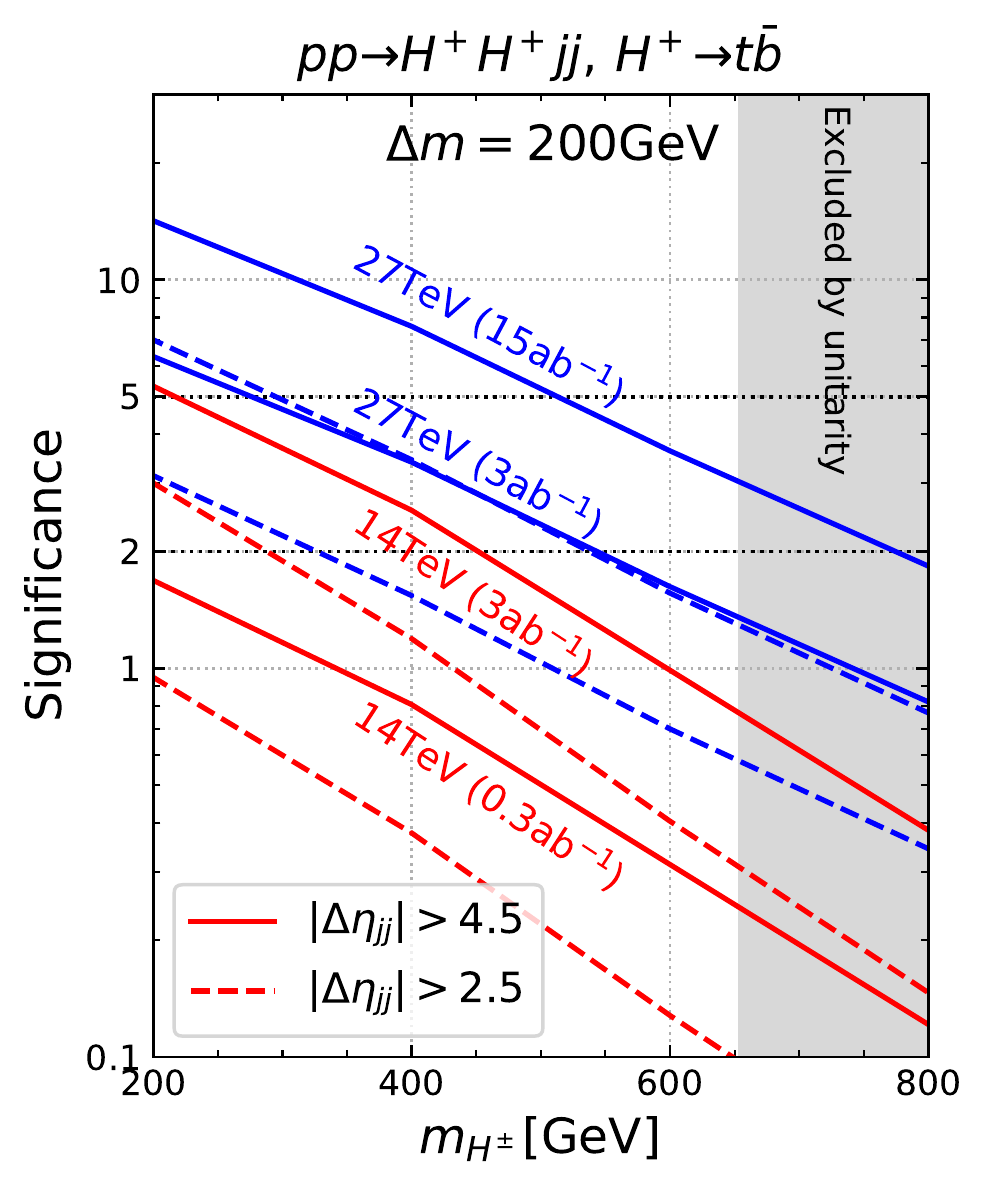}%
  \caption{
Signal significance of $pp\to H^+H^+jj$ for $H^+\to\tau^+\nu$ (left) and $H^+\to t\bar b$ (right). 
 }
\label{fig:sig}
\end{figure*}

Second, signature for the signal $pp\to H^\pm H^\pm jj$ with $\ H^\pm\to tb$ is 
a same-sign top-quark pair plus two forward jets.
The main SM background would be four top-quark ($tt\bar t\bar t$) production, 
whose cross section was recently measured 
at the LHC~\cite{Sirunyan:2017roi}.
To identify the electric charge of the top quarks, we consider the leptonic decay of top quarks
and the hadronic decay of anti-top quarks.
We estimate the event numbers as
\begin{align}
  s=& L\; \sigma^{\rm VBF}_{pp\to H^+ H^+jj}  (B_{H^+\to t\bar b})^2 
  (B_{t\to b\ell^+ \nu})^2  \epsilon_{\rm sel}^t (\epsilon_{b})^{2}, \notag\\  
  b=&L\; \sigma_{pp\to tt\bar t\bar t} 
     (B_{t\to b\ell^+ \nu})^2     (B_{\bar t\to \bar bjj})^2
     \epsilon_{\rm sel}^t  (\epsilon_{b})^{2},  \notag
\end{align}  
where $\ell=e$ or $\mu$.
The four-top cross section at $\sqrt{s}=14$ (27)~TeV is estimated at NLO as 16 (144)~fb~\cite{Frederix:2017wme}.  
$B_{H^+ \to t \bar{b}}=1$ is taken.   
In addition to the VBF baseline selection for jets, 
we require two same-sign leptons with
$p_T^\ell>20$~GeV and $|\eta^\ell|<2.5$.
Moreover, to suppress other backgrounds such as from multi-boson production, 
we also require the scalar sum of the transverse momenta of all jets to be $H_T>300$~GeV and 
at least two $b$-tagged jets with the tagging efficiency $\epsilon_b=0.65$~\cite{Sirunyan:2017roi}.
The selection efficiency $\epsilon_{\rm sel}^t$ for the signal is 0.56--0.91 (0.49--0.88) 
for $m_{H^\pm}=200$--1000~GeV with $|\Delta\eta_{jj}|>2.5$ (4.5) at $\sqrt{s}=14$~TeV,
while that for the background is 0.031 (0.004).

Finally, in Fig.~\ref{fig:sig}, 
we show signal significances of $pp\to H^+H^+jj$ with the $H^+\to\tau^+\nu$ decay (left) and those with the $H^+\to t\bar b$ decay (right)
as a function of $m_{H^\pm}$,  where $\Delta m=200$~GeV is taken. 
The significances with the VBF baseline selection plus a larger rapidity separation cut are shown by dashed and solid lines, respectively,
at the 14TeV LHC with $L=0.3$ and $3$ ab$^{-1}$ as well as at a future collider ($\sqrt{s}=27$ TeV) with $L=3$ and $15$ ab$^{-1}$. 
For simplicity, we use the same kinematical cuts, and assume the same efficiencies 
of the hadronic tau and the $b$-jet identifications for the collider with $\sqrt{s}=27$~TeV.
The shaded regions are excluded by unitarity bounds~\cite{Kanemura:1993hm}.  
The sensitivities can be significant for smaller $m_{H^\pm}$.

%[Conclusions]

In summary, we discussed the new interesting process 
$pp \to W^{\pm\ast} W^{\pm\ast} jj \to H^\pm H^\pm jj$ in THDMs,  
which is timely and useful to explore the global symmetry structure of the Higgs potential.   
We evaluated the sensitivities for $H^\pm\to\tau\nu$ and $H^\pm\to tb$ decays  
at the LHC and future higher-energy colliders. 
We found that the process can be feasible. 
The details are given elsewhere~\cite{AKM2}. 

%\section*{Acknowledgements}

The authors would like to thank H. E. Haber for useful discussions.
This work was supported,  in part, by MEXT Grants No.~16H06492 and No.~18H04587, 
and JSPS Grant No.~18K03648.

%\bibliography{bib_higgs}
%bibliographystyle{apsrev4-1} 

%

\end{document}